\documentclass[preprint,tightenlines,superscriptaddress,prd,nofootinbib,eqsecnum,showpacs]{revtex4}

\usepackage{amssymb}\usepackage{graphicx}
\usepackage{enumerate}
\def\dd{{\rm d}}\def\half{{\textstyle\frac12}}
\def\_#1{^{}_{\scriptscriptstyle\rm #1}}
\def\beq{\begin{equation}}\def\eeq{\end{equation}}
\def\bea{\begin{eqnarray}}\def\eea{\end{eqnarray}}

\begin{document}

\title{Effect of Accelerated Global Expansion on Bending of Light}

\author{Mir Emad Aghili}\email{maghili@go.olemiss.edu}
\affiliation{Department of Physics and Astronomy,
The University of Mississippi, University, MS 38677}

\author{Brett Bolen}\email{bolenbr@gvsu.edu}
\affiliation{Department of Physics,
Grand Valley State University, Allendale, MI 49504\\ \kern10pt}

\author{Luca Bombelli}\email{luca@phy.olemiss.edu}
\affiliation{Department of Physics and Astronomy,
The University of Mississippi, University, MS 38677}

\date{May 25, 2016}

\begin{abstract}\noindent In 2007 Rindler and Ishak showed that, contrary to previous claims, the value of the cosmological constant does have an effect on light deflection by a gravitating object in an expanding universe. In their work they considered a Schwarzschild-de~Sitter (SdS) spacetime, which has a constant asymptotic expansion rate $H_0$. A model with a time-dependent $H(t)$ was studied by Kantowski et al., who consider in their 2010 paper a ``Swiss-cheese" model of a Friedmann-Lema\^itre-Robertson-Walker (FLRW) spacetime with an embedded SdS bubble. In this paper, we generalize the Rindler and Ishak model to time-varying $H(t)$ in another way, by considering light bending in a McVittie metric representing a gravitating object in a FLRW cosmological background. We carry out numerical simulations of the propagation of null geodesics in different McVittie spacetimes, in which we keep the values of the distances from the observer to the lensing object and to the source fixed, and vary the form of $H(t)$.
\end{abstract}
\maketitle

\bigskip
\section{Introduction}

\noindent The effect of the cosmological expansion on the bending of light rays and gravitational lensing in situations where the expansion is driven by a cosmological constant has been studied for more than 30 years, since Islam's 1983 paper \cite{Islam} on light trajectories in Schwarzschild-de~Sitter (SdS) spacetime, but the issue has received an increased amount of attention in the past 15 or so years with observations leading to the conclusion that the global rate of expansion is accelerating. Among the recent references on the subject, we refer the reader to the papers by Ishak and Rindler \cite{Rindler1,Rindler2}, and to the comprehensive discussion by Lebedev and Lake \cite{KLake}, which also contain citations to the rest of the literature. In this paper we will examine light bending in a more general setting, but still representing a nonrotating gravitating object in an asymptotically homogeneous and isotropic background.

Let us introduce some relevant concepts and model spacetimes. A spatially flat, homogeneous, isotropic cosmological model is described by a Friedmann-Lema\^itre-Robertson-Walker (FLRW) metric of the form $\dd s^2 = -\dd t^2 + a(t)^2\,\dd\vec x^{\,2}$, where $\dd\vec x^{\,2}$ is the line element for Euclidean 3-space, in the time gauge in which $t$ is proper time along a comoving worldline, with spatial scale factor $a(t)$. The Hubble parameter $H(t):= \dot a/a$ is usually referred to as the ``expansion rate"; an accelerating expansion rate corresponds to a situation in which $\ddot a\ne 0$, and is often quantified by the ``deceleration parameter" $q:= -\ddot{a}a/\dot{a}^2 \equiv -(1+\dot H/H^2)$.

In the standard cosmological model, results of observations such as the redshift-luminosity relationship for supernovae are interpreted as indicating that $q < 0$. The physical reason for this acceleration is not yet well understood, but one simple possibility is the presence of a non-vanishing cosmological constant, and most of the literature on gravitational light bending and the cosmological expansion has focused so far on this possibility. A cosmological constant $\Lambda$ corresponds to constant values $H = H_0 = \sqrt{\Lambda/3}$ and $q = -1$, and the model used to study its effects on light bending is the Schwarzschild-de Sitter spacetime, which in the common ``static" coordinates takes the form of the Kottler metric \cite{Kottler},
\beq
\dd s^2 = -\left(1-\frac{2m}r-\frac{\Lambda r^2}3\right)\dd t^2
+ \left(1-\frac{2m}r-\frac{\Lambda r^2}3\right)^{\!-1}\dd r^2
+ r^2\,\dd\Omega^2\;. \label{SchwarzschilddS}
\eeq
This line element approaches that of the Schwarzschild metric for a black hole of mass $m$ located at the center of the coordinate system for small $r$ (or $\Lambda = 0$), and the de~Sitter metric with cosmological constant $\Lambda$, written in static coordinates, for large $r$ (or $m = 0$).

In the presence of matter fields, however, or if the accelerated expansion is due to any reason other than a cosmological constant, we do not expect $H(t)$ to be constant. Indeed, in the standard $\Lambda$CDM cosmological model the time dependence of $H(t)$ is governed by both $\Lambda$ and different types of matter, and observationally there are indications that $H$ has been varying in time (see, for example, Refs.\ \cite{PanSTARRS,Odderskov}). Our goal is to extend the work on gravitational light bending at cosmological scales to the case of a time-dependent expansion rate $H(t)$ and study the effects of a non-zero $\dot H(t)$.

A model with time-dependent Hubble parameter was analyzed in an 2010 article by Kantowski et al.\ \cite{Kantowski}, who used a ``Swiss cheese" model to examine the bending of light in a cosmological setting.  In that paper, the authors consider a FLRW cosmology where a spherical region has been replaced by a bubble inside which the metric is of the SdS type. The values of the bubble radius $r_{\rm b}$, black-hole mass $m$ and cosmological constant $\Lambda$ are subject to constraints from matching conditions at the bubble boundary; in particular, $r_{\rm b}$ scales in time the same way $a(t)$ does.  They can then perturbatively calculate the lightlike geodesics up to orders $(m/r_0)^2$ and $(m/r_0) \Lambda r_0^2$, where $r_0$ is the distance of closest approach to the central black hole. Using realistic parameter values in a $\Lambda$CDM model, the authors found that the combined effects of the additional $\Lambda$ and matter terms on the bending angles could be as large as a few percent.

Our approach will be to model a gravitating object in an expanding universe with non-constant $H(t)$ by a McVittie metric, and examine the behavior of null geodesics in it. In this model we avoid the need for ``gluing" together two different solutions of Einstein's equation and we can prescribe an arbitrary expansion rate; the price we pay is having an unphysical equation of state for the fluid matter, and the need to integrate the geodesic equation numerically; in this sense, our results are complementary to those of Kantowski et al. The McVittie metric has been recently used in work on the effect of cosmological expansion on gravitational lensing \cite{Piattella}, but in that paper $H$ was taken to be constant and, as we will see in the next section, the spacetime is then equivalent to Schwarzschild-de Sitter. In Sec.\ \ref{McVittiereview} we review the McVittie metric, written in two convenient coordinate systems. In Sec.\ \ref{SdSbending} we summarize previous results on the bending of light in the SdS spacetime. Sec.\ \ref{McVittiebending} contains our results on light bending in McVittie spacetimes with varying $H(t)$, and we finish with some conclusions in Sec.\ \ref{Conclusions}. We use units in which $c = G = 1$.

\section{McVittie Metric}
\label{McVittiereview}

\noindent A metric that can be thought of as representing a Schwarzschild black hole embedded in a fluid-filled FLRW spacetime was first derived in 1933 by G.C. McVittie \cite{McVittie}. Surprisingly, despite it being a rather old solution of the Einstein equation, its exact interpretation has been somewhat controversial, and papers have appeared questioning the black-hole interpretation of the metric \cite{Nolan1,Nolan2}; recent work \cite{Kaloper,Faraoni} has shown, however, that this metric does indeed model a black hole embedded in an expanding cosmology, at least when the background spacetime is spatially flat and the Hubble parameter satisfies $H(t) > 0$. We should point out though that the McVittie metric is not considered as representing a generic solution of this type. Physically, in this situation one would expect the energy density to be inhomogeneous, due to the expansion of the universe at large scales and the attraction of the black hole at small scales, which might cause the black hole to accrete matter as in a cosmological analog of the Vaidya metric \cite{Vaidya,Poisson}. As Kaloper et al.\ argue, this is not the case; the equation relating the energy density to $H(t)$ is exactly the same as for a FLRW cosmology, while the inhomogeneity appears in the expression for the pressure \cite{Kaloper}.

We use the McVittie metric in this work because it allows us to freely specify the function $H(t)$ in an asymptotically FLRW model with a localized gravitating mass $m$; since we will consider the light ray as interacting only with the geometry, the form of the matter equation of state will not be directly relevant for us. Of course, that equation of state does affect the local spacetime geometry, but the fact that our local geometry does not have the sharp bubble boundaries of the Swiss-cheese universe makes it arguably more realistic. The McVittie line element can be written down in the diagonal form 
\beq
\dd s^2= - \left(\frac{1-\mu}{1+\mu}\right)^{\!2} \dd t^2
+ (1+\mu)^4\, a^2(t)\, \dd\vec X^{\,2}\;,\qquad
\mu:= \frac{m}{2\,a(t)R}\;, \label{comoving}
\eeq
where $a(t)$ is the asymptotic scale factor; these coordinates are an inhomogeneous-space version of comoving coordinates, in the sense that at large values of $|\vec X|$ the line element tends to that of a FLRW spacetime with scale factor $a(t)$, expressed in comoving coordinates.

Several authors \cite{Kaloper, Bolen, Nandra} have introduced a coordinate transformation $(t,\vec X) \mapsto (t,\vec x)$ to ``Painlev\'e-Gullstrand type" coordinates defined by
\beq
r = (1+\mu)^2 a(t)\,R \label{staticR}
\eeq
(the coordinates $t$, $\theta$ and $\phi$ are unchanged), in which the line element takes the form
\beq
\dd s^2= -\Big(f(r)- H^2(t)\, r^2\Big)\, \dd t^2- \frac{2H(t)\,r}{\sqrt{f(r)}}\, \dd r\,\dd t + \frac{\dd r^2}{f(r)} + r^2\, \dd\Omega^2 \;, \label{static}
\eeq
with $f(r):= 1-2m/r$ and again $H(t) = \dot{a}/a$. Thus, $r$ is an area radial coordinate, characterized by the fact that the surface area of a 2-sphere of constant $r$ is $4\pi r^2$, and the function $H(t)$ represents the asymptotic expansion rate in comoving coordinates, so we still call it the ``Hubble parameter". Note that in the ADM-type terminology for the 3+1 split of a spacetime metric, the spatial part of this metric has no time dependence and has the same form as in Schwarzschild spacetime, while the cosmological expansion terms have been moved into the shift vector, with non-vanishing component $N_r = -H(t)\,r/\sqrt{f(r)}$.

For us, it is importat to note that the SdS spacetime with cosmological constant $\Lambda$ is a special case of the McVittie metric with constant $H(t) = H_0$ and $\Lambda = 3H_0^2$, as one can see by defining the time coordinate transformation
\beq
\dd t = \dd\bar t -\frac{1}{1-\frac{2m}{r} -\frac{1}{3} \Lambda r^2}
\,\sqrt{\frac{\Lambda r^2/3}{1-2m/r}}\, \dd r\;. \label{newt}
\eeq
Using $\bar t$ as time coordinate, the McVittie metric takes the traditional Kottler form of Eq.\ (\ref{SchwarzschilddS}) (see, e.g., Ref.\ \cite{Nandra}).

In the literature on light deflection in SdS spacetimes authors use the Kottler, diagonal form of the metric, because of its greater simplicity. If $H(t)$ is not constant, however, a coordinate transformation of the type (\ref{newt}) does not yield a diagonal, ``generalized Kottler" metric with ``time-varying $\Lambda$", and we will use instead the form (\ref{static}). We will call (\ref{static}) the (spatially) ``static" form of the McVittie metric, and the corresponding observers ``static", as opposed to the ``comoving" ones of the line element (\ref{comoving}), in which the spatial metric does not have a static form even when $H(t)$ is constant.

\section{Null Geodesics and Light Bending in SDS Spacetimes}
\label{SdSbending}

\noindent In this section we will review some of the definitions and results in the literature on null geodesics in SdS spacetime (see, for example, the papers by Ishak and Rindler \cite{Rindler1,Rindler2,Ishak} and Lebedev \& Lake \cite{KLake}). In view of the fact that we will later generalize those results, however, we will replace $\Lambda$ by $3H_0^2$, where $H_0$ is the constant value of $H(t)$, use the line element (\ref{static}), and add some remarks motivated by the more general situation.

The {\it whole\/} McVittie line element is now time-independent, and we can use the static and rotational Killing vector fields $\xi = \partial/\partial t$ and $\eta = \partial/\partial\phi$ to define two conserved quantities along a geodesic $x^\mu(\lambda)$ with tangent vector $K^\mu:= \dd x^\mu/\dd\lambda$,
\beq
e:= -g_{\alpha\beta}^{}\, \xi^\alpha K^\beta
= [f(r)-H_0^2 r^2]\, K^t + \frac{H_0\,r}{\sqrt{f(r)}}\, K^r \;,\qquad
\ell:= g_{\alpha\beta}^{}\, \eta^\alpha K^\beta = r^2\,K^\phi\;,
\eeq
for a geodesic in the equatorial coordinate plane. In other words, the geodesic satisfies
\beq
K^\phi = {\ell\over r^2}\;,\qquad
K^t = {e\over f(r)-H_0^2\,r^2} - {H_0r\over\sqrt{f(r)}\,[f(r)-H_0^2\,r^2]}\,K^r\;.
\eeq
Using these equations, the conditions for a null geodesic with $g_{\mu\nu}\,K^\mu K^\nu = 0$ can be reduced to a single differential equation, which in terms of the conserved quantities becomes
\beq
\frac{1}{\ell^2} \left( \frac{\dd r}{\dd \lambda}\right)^{\!2}
+ \frac{1}{r^2}\, (f(r) - H_0^2\, r^2) = \frac{1}{b^2} \;, \label{schwarzdrdlambda}
\eeq
where we have defined $b:= \ell/e$, the impact parameter. Note that, as expected, if $H_0 \to 0$ this is the same equation one gets for a null geodesic in Schwarzschild spacetime (see, for example, Ref.\ \cite{Hartle} or most of the references on light bending in SdS spacetime).

The shape of the light trajectory, in the sense of the relationship between the coordinates $r$ and $\phi$, can be obtained using the expression for $\ell$ in (\ref{schwarzdrdlambda}), and one finds that 
\beq
\frac{\dd\phi}{\dd r} = \pm \frac{1}{r^2}
\left[ \frac{1}{b^2} + H_0^2 - \left(1- \frac{2m}{r}\right)\right]^{-1/2}.
\label{schwdphidr}
\eeq
Notice that, since these equations do not involve $t$ or the geometry of a constant-$t$ hypersurface, we would have obtained the same relationship between $r$ and $\phi$ using the form (\ref{SchwarzschilddS}) for the metric. Eq.\ (\ref{schwdphidr}) encodes some aspects of the effect a cosmological constant has on the formal description of light bending near a nonrotating gravitating mass, but one should be careful when discussing any measurable effects on observations. A controversy in this sense arose around the interpretation of (\ref{schwdphidr}), because although the equation does depend on $\Lambda$, it is of the same form as the one obtained in Schwarzschild spacetime if one replaces the impact parameter $b$ in the latter by an ``effective impact parameter" $B$ defined by
\beq
\frac{1}{B^2}:= \frac{1}{b^2} + H_0^2\;, \label{impact}
\eeq
and one might conclude that the cosmological constant does not affect observations because it is simply ``absorbed" into the impact parameter, which is not directly measurable. This aspect has been extensively discussed starting with the original paper by Islam \cite{Islam}. However, the more recent literature has made it clear that the relationship between $r$ and $\phi$ only tells us part of the story on how the deflection angle depends on $\Lambda$, for various reasons.

\begin{figure}
\includegraphics[angle=0,scale=1.15]{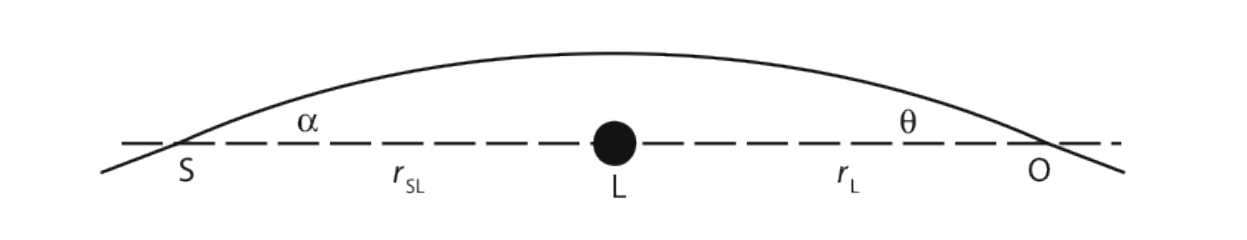}
\caption{\label{bendingfigure}Definition of the main angles and static-coordinate distances used in the calculation of the light bending angle. The lensing object $L$ is located a distance $r\_L$ and the light source at a distance $r\_S$ from the observer $O$. The full bending angle is $\alpha + \theta$ but only $\theta$ is measurable by the observer. In SdS spacetime the point of closest approach is at $\phi = \pi/2$ and the figure is symmetric about the vertical axis; in general, this is no longer the case in McVittie spacetime.}
\end{figure}

One reason is that, as pointed out by Ishak and Rindler \cite{Rindler1,Rindler2}, from (\ref{schwdphidr}) alone and without knowing the spatial metric, one can only obtain the value that an angle would have in flat space. Consider for example the angle $\theta$ in Fig.\ \ref{bendingfigure}. This is not the full bending angle, but as the following argument illustrates it is relevant for the calculation of the latter. In Schwarzschild spacetime one can imagine placing the source and observer at infinity and calculate a bending angle that does not refer to specific locations for them and depends only on the impact parameter $b$, in addition to the lens mass $m$. Because of the curved geometry of the constant-$t$ surfaces, this is not an option in SdS spacetime \cite{Rindler1,Rindler2}. The simplest alternative is to set up the coordinates so that the point of closest approach is at $\phi = \pi/2$ and use as measure of total bending the sum $\alpha + \theta$ of the angles the light ray makes with the half-lines $\phi = \pi$ and $\phi = 0$ at the intersection points, labeled $S$ and $O$ in Fig.\ \ref{bendingfigure}. Because the spacetime is static and spherically symmetric, in this situation if the radial positions of $S$ and $O$ are equal, $r\_{SL} = r\_L$, then so are the angles, $\alpha = \theta$. The ``Euclidean" value $\theta\_E = \tan^{-1}(r\dd\phi/\dd r)_{_O}$ of half of the bending angle can then be easily calculated; neglecting quadratic terms in $m$, one finds the well-known result (see, e.g., Refs.\ \cite{Rindler1,Rindler2})
\beq
\theta\_E \approx \frac{2m}{B} \approx \frac{2m}{b} + mbH_0^2
= \frac{2m}{b} + \frac{mb\Lambda}{3}\;. \label{EuclideanAngle}
\eeq
However, a more physically meaningful, covariant value for the angle between two lines takes into account the actual metric on a constant-$t$ spatial hypersurface in SdS spacetime. As first derived by Ishak \& Rindler \cite{Rindler1,Rindler2}, this measurable value $\theta\_M$ of the angle $\theta$, to leading order in $\Lambda$, is given by an expression with a different $H_0$-dependent term,
\beq
\theta\_M \approx \frac{2m}{b} - \frac{\Lambda b^3}{12m}
= \frac{2m}{b} - \frac{H_0^2 b^3}{4m}\;. \label{alphac}
\eeq

Secondly, when discussing observational consequences of $H_0$ for light bending we need to specify exactly what question we are asking. For example, if we said that Eq.\ (\ref{EuclideanAngle}) implies that the Euclidean bending angle $\theta\_E$ depends on $H_0^{}$ through the effective parameter $B$, we would be assuming that the comparison is made between situations in which the impact parameter $b$ is held constant. However, because $b$ is not directly measurable, it may be more meaningful to specify in some other way which situations we are comparing when we change the value of $H_0$. We take the point of view that, for each value of $H_0^{}$, the geodesic we use to find the angle $\theta$ needs to be uniquely determined by a fixed set of values for parameters that can in principle be measured by an observer. As we will see in the next section, our choice will lead to a different conclusion for the dependence of $\theta\_E$ on $H_0^{}$.

The third thing to keep in mind is that values of angles are observer-dependent. Thus, the angle $\theta\_M$ in (\ref{alphac}) is the one that would be measured by a ``static observer", one along whose worldline $r$ and the angular coordinates are constant, but one may want to know what angle would be measured by a ``comoving observer", one along whose worldline $R$ rather than $r$ is constant. This point is discussed in detail by Lebedev \& Lake \cite{KLake}. A general expression for the angle measured by an observer with 4-velocity $U^\mu$ can be calculated from the dot product between the projections orthogonal to $U^\mu$ of the vector $K^\mu$ tangent to the deflected null geodesic and a radial null vector $W^\mu$ (see Eq.\ 80 in Ref.\ \cite{KLake}),
\beq
\cos\theta\_M = \frac{K\cdot W}{(U\cdot K)(U\cdot W)}+1\;.\label{thetao}
\eeq
While we will not go into the details of how this is used to derive explicit expressions for the angles measured by different observers, let us summarize some results we can compare with our numerical ones for McVittie spacetime in the next section. From (\ref{thetao}) one can derive a relationship between the Euclidean angle $\theta\_E$ and the angle $\theta\_M$ measured by a static observer in SdS spacetime \cite{KLake},
\beq
\tan\theta\_M = \sqrt{f(r)-H_0^2\, r^2}\, \tan\theta\_E \;. \label{tantheta}
\eeq
On the other hand, from Eq.\ (\ref{schwarzdrdlambda}) for null geodesics one obtains the well-known first-order lens equation, valid for the general case of a source at a distance $y$ from the lens-observer axis (see, for example, Refs.\ \cite{Rindler1,Rindler2,KLake}). The result, using the notation $r\_S = r\_{SL} + r\_L$, is
\beq
y = r\_S\, \theta\_E - \frac{4m\,r\_{SL}}{r\_L\,\theta\_E}\;. \label{y}
\eeq
When the source is on the axis, $y = 0$, if we solve (\ref{y}) for the Euclidean angle we get
\beq
\theta\_E = \sqrt{\frac{4m\,r\_{SL}}{r\_L\,r\_S}}\;; \label{thetaE}
\eeq
this $\theta\_E$ can in turn be plugged into (\ref{tantheta}) and, using the small-angle approximation, for the static-observer measurable angle we get
\beq
\theta\_M = \sqrt{\left({1-\frac{2m}{r\_L}-H_0^2\, r_{\scriptscriptstyle\rm L}^2}\right)
\frac{4m\,r\_{SL}}{r\_L\,r\_S}}\;. \label{thetastatic}
\eeq
An expression for the comoving-observer measurable angle, which is physically related to the static-observer one by an appropriate aberration factor, was obtained in Refs.\ \cite{Rindler1,Rindler2,KLake}. For more comments on formulating physically relevant questions about the effect of $\Lambda$ on the bending angle in SdS spacetime, we refer to the work of Ishak and Rindler \cite{Rindler1,Rindler2} or the more recent papers by Lebedev and Lake \cite{KLake} and by Hammad \cite{Hammad}.

\section{Null geodesics in McVittie spacetime}
\label{McVittiebending}

\noindent Let us now consider light bending in a McVittie spacetime with non-constant expansion rate $H(t)$. In the coordinates of the ``static" line element (\ref{static}), a light ray in this spacetime obeys the null condition
\beq
\left[f(r) - r^2 H(t)^2 \right] (K^t)^2 + \frac{2rH(t)}{\sqrt{f(r)}}\, (K^r K^t)
- \frac{1}{f(r)}\, (K^r)^2 - \frac{\ell^2}{r^2} = 0\;, \label{McVittieNull}
\eeq
where as before the angular momentum $\ell = r^2\,K^\phi$ is conserved. Unlike in the earlier SdS case, however, there is no second conserved quantity $e$, because the spacetime has no timelike Killing vector field in general. Therefore, to find the deflected light ray and determine its tangent vector at the observer's location we must actually solve at least one component of the null geodesic equation in addition to (\ref{McVittieNull}), as opposed to just using first integrals.

Using the explicit form of the connection coefficients $\Gamma^\mu{}_{\alpha\beta}$
for the McVittie metric (\ref{static}), the relevant components of the geodesic equation
$\dd K^\mu/\dd\lambda + \Gamma^\mu{}_{\alpha\beta}\, K^\alpha K^\beta = 0$ become
\bea
& &\frac{\dd K^t}{\dd\lambda} + \left(\frac{rH(t)
\left[2 rH(t)^2-f'(r)\right]}{2\sqrt{f(r)}} \right) (K^t)^2 +
\frac{f'(r)-2 r H(t)^2}{ f(r)}\, K^r K^t \nonumber \\
& &\kern23pt+\ \frac{H(t)}{f(r)^{3/2}}\, (K^r)^2
+ \frac{H(t)\, \ell^2}{\sqrt{f(r)}\, r^2} = 0 \label{utgeodesic} \\
\noalign{\smallskip}
& &\frac{\dd K^r}{\dd\lambda} 
+ \frac{2 r H(t)^2- f'(r)}{2 f(r)} \, (K^r)^2 
+ \frac{r H(t)}{ \sqrt{f(r)}}\, (f'(r)-2 r H(t)^2)\, K^tK^r
+ \frac{\ell^2}{r^3}\left( r^2 H^2(t)- f(r)\right) \nonumber \\
& &\kern 24pt +\ \frac{1}{2} \left( (f(r)-r^2 H^2(t))(f'(r) -2 r H^2(t))
-2 r \sqrt{f(r)} H'(t) \right) (K^t)^2 = 0\;, \label{urgeodesic}
\eea
and our goal is to solve Eqs.\ (\ref{utgeodesic}) and (\ref{urgeodesic}) for $K^\mu = \dd x^\mu/\dd\lambda$, with initial values at the source satisfying the condition (\ref{McVittieNull}). We do not know how to solve those equations analytically, so we integrated them numerically instead, using the standard fourth-order Runge-Kutta method with approximately $10^{6}$ steps, and a variable integration step which according to standard theory \cite{NumRec} leads to cumulative relative errors due to the Runge-Kutta approximation at most of order $10^{-13}$ for our variables. A further accuracy check was performed by comparing the final value of $K^t$ produced by each simulation to that which is demanded by the null normalization condition $K^\mu K_\mu =0$ using the final values for the other variables; we found relative differences of order $10^{-12}$, indicating that in practice those errors were dominated by roundoff errors.

Before describing the results of our numerical simulations, we need to discuss how we will determine the effect of the time dependence of $H(t)$ on the light bending angle. We will be comparing with each other McVittie metrics with the same mass $m$ and the same value of the Hubble parameter  $H(t)$ at the time $t_0$ when the null geodesic leaves the source, but different $\dot H(t)$. Treating $H(t)$ as a slowly varying function over the relevant times, we will parameterize it simply by giving the values of $H_0 = H(t_0)$ and $A = \dot H(t_0)$, or
\beq
H(t) = H_0 + A\,(t-t_0)\;. \label{linear}
\eeq
This parametrization is a generic one for small $A$, and it allows us to study the effect of different time rates of change of $H(t)$ on the bending angle using a range of values for $A$ motivated by data from cosmological observations, rather than choosing a value tied to a specific cosmological model.

Comparing predictions for the bending angle as $A$ varies, however, raises a conceptual issue. Because McVittie metrics with different values of $A$ are to be thought as entirely different spacetime manifolds, a key point when comparing them is to formulate a criterion for determining which null geodesic in each spacetime is to be used for the comparison. One possibility, mentioned when we discussed SdS metrics, might have been to use geodesics with the same impact parameter $b$. We take instead the point of view that a meaningful criterion should be based on quantities an observer has access to.

For simplicity, in this paper we will assume that observers are able to measure the distances to the source and lens (which to leading order in $m$ we identify with the values of $r\_S$ and $r\_L$), in addition to the angle $\theta\_M$, as well as the fact that the source and lens are aligned, and look for how $\theta\_M$ varies with $A$ when $m$, $H_0$, $r\_S$ and $r\_L$ are kept constant. In a McVittie spacetime, fixing a set of values for those parameters determines uniquely a pair of null geodesics, one on each side of the lens, and for definiteness we will choose the clockwise one, as in Fig.\ \ref{bendingfigure}. Considering $r\_S$ and $r\_L$ as directly measurable is overly simplistic, and considering only cases in which the source, lensing object and observer are aligned limits the generality of the results. We leave it as a goal for future work to remove those limitations, but we should also point out that aligned configurations are the physically most relevant ones; although light bending certainly occurs in more general settings, it is much more likely to be noticed observationally when all objects are nearly aligned, or $y = 0$ in Eq.\ (\ref{y}).

\begin{figure}
\includegraphics[scale=0.35]{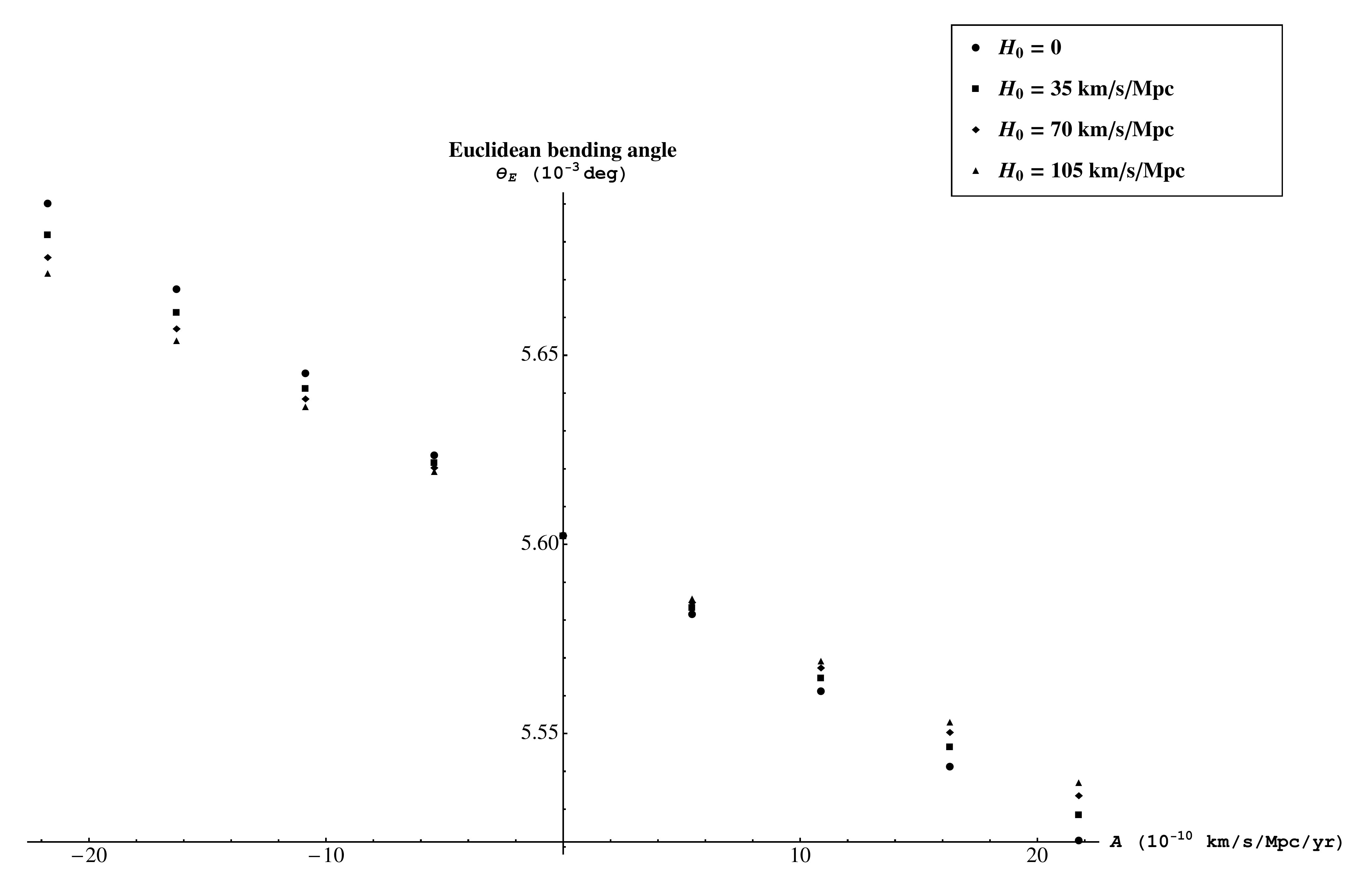}
\caption{\label{angleE} A plot of the Euclidean angle $\theta\_E$ for an observer at fixed distances $r\_S = 2$ Gpc from the source and $r\_L = 1$ Gpc from the lensing object (in static coordinates) vs.\ the acceleration parameter $A$, for various values of $H_0^{}$. For $A = 0$ the bending angle is independent of $H_0^{}$, to a good approximation, and its value agrees with Eq.\ (\ref{thetaE}).}
\end{figure}

\begin{figure}
\includegraphics[scale=0.35]{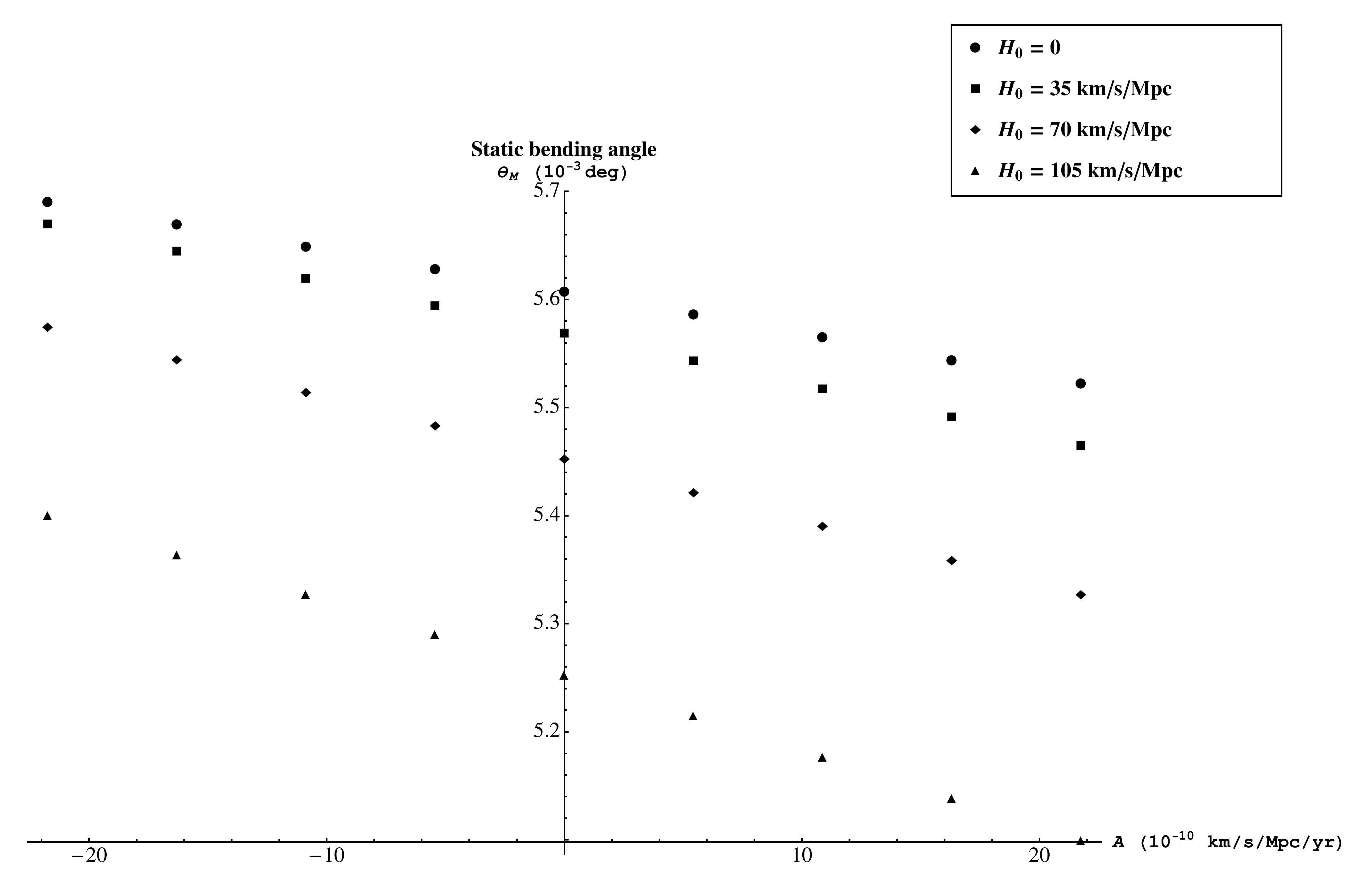}
\caption{\label{angleS} A plot of the angle $\theta\_M$ measured by a static observer at fixed distances $r\_S = 2$ Gpc from the source and $r\_L = 1$ Gpc from the lensing object (in static coordinates) vs.\ the acceleration parameter $A$, for various values of $H_0^{}$. For each value of $A$, the arrival angle decreases with $H_0$. The direction of change with respect to $H_0$ is the same as the direction of change with $A$.}
\end{figure}

\begin{figure}
\includegraphics[scale=0.35]{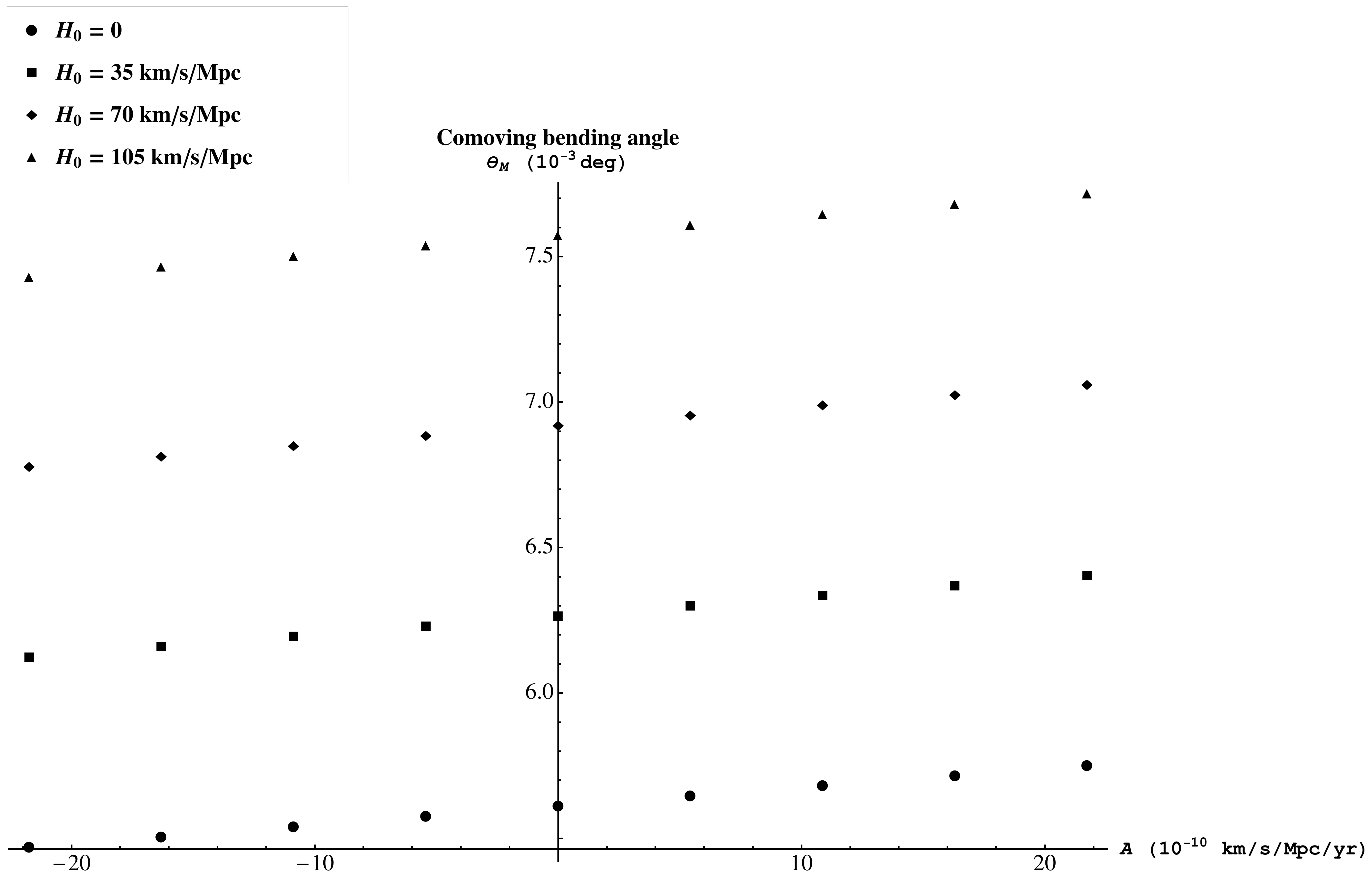}
\caption{\label{angleC} A plot of the angle $\theta\_M$ measured by a comoving observer at fixed distances $r\_S = 2$ Gpc from the source and $r\_L = 1$ Gpc from the lensing object (in static coordinates) vs.\ the acceleration parameter $A$, for various values of $H_0^{}$. For each value of $A$, the arrival angle increases with $H_0$. The direction of change with respect to $H_0$ is the same as the direction of change with $A$.}
\end{figure}

When performing the simulations, we start each light ray from the coordinate location $r = r\_{SL} = r\_S - r\_L$, $\phi = \pi$ at $t = t_0$, and aim it in a trial direction $\alpha$ towards decreasing values of $\phi$. The geodesic equations (\ref{utgeodesic}) and (\ref{urgeodesic}) are then integrated numerically until the light ray reaches $\phi = 0$, where the value of $r$ is checked against the chosen $r\_L$. If we find that $r$ at $\phi = 0$ is smaller (greater) than the desired value, the simulation is repeated using a larger (smaller) value for $\alpha$ until the final $r$ equals $r\_L$, at which point the components of $K^\mu$ are recorded. The relative tolerance when the value of $r$ is matched with $r\_L$ is $10^{-6}$; this is the largest source of errors in our results for the bending angles, and it is small enough that the corresponding error bars do not show up in Figs.\ \ref{angleE}--\ref{angleC}. Notice that when $H(t)$ is time-dependent the spatial projection of a deflected null geodesic is not symmetric about the point of closest approach, and in a sketch of the trajectory of the light ray similar to the SdS one in Fig.\ \ref{bendingfigure}, even with $r\_{SL} = r\_L$, the angles $\alpha$ and $\theta$ will not be equal and the point of closest approach will not be at $\phi = \pi/2$.

Once the components of $K^\mu$ at the location of the observer are known, we use (\ref{thetao}) to calculate the angle $\theta\_M$ measured by an observer with 4-velocity $U^\mu$. For static and comoving observers the static-coordinate components of the 4-velocities are
\beq
U^\mu_{\rm stat} = \Bigg(\frac{1}{\sqrt{f(r)-H^2(t)\,r^2}},\,0,\,0,\,0\Bigg)\;,\quad
U^\mu_{\rm com}
= \Bigg(\frac{1}{\sqrt{f(r)}},H(t)\,r,\,0,\,0\Bigg)\;,\;\label{4velo}
\eeq
respectively, and finding a null radial vector is simple; we choose
\beq
W^\mu = \Big(1,\sqrt{f(r)}\,\Big[\sqrt{f(r)} + Hr\Big],0,0\Big)\;.\label{radnull}
\eeq

To specify the metric we choose a single value $m = 10^{14} \, M_{\rm sun}^{}$ for the mass of the lensing object, considered to be representative of situations in which a light ray traveling over cosmological distances is deflected by a galaxy cluster. For $H_0$ we choose three values in the range from 0 to 70 km/s/Mpc (from no expansion to approximately the currently accepted value). To estimate a realistic range of values for $A$ we use the fact that $q = \half\,(1+3w)$, where $w$ is the cosmological equation of state parameter, and we take values consistent with current estimates to be approximately in the range from 1.05 to 1.25 \cite{PanSTARRS}. Based on this and the relationship $\dot H = -(1+q) H_0{}^2$, we choose the range from $-1.0\times10^{-9}$ to $+1.0\times10^{-9}$ km/s/Mpc/yr for the values for $A$. To specify the null geodesic, for the coordinate values of the source-lens and lens-observer distances we choose $r\_{SL} = r\_L = 1$ Gpc, considered as representative of typical values in a cosmological lensing situation, with a tolerance of 0.1 Mpc in the value of $r$ for the intersection of the geodesic with $\phi = 0$.

The results of our simulations are shown in Figs.\ \ref{angleE}--\ref{angleC}. For comparison with what we said in Sec.\ \ref{SdSbending} about SdS spacetime, Fig.\ \ref{angleE} shows the Euclidean angle $\theta\_E = \tan^{-1}(r\,K^\phi/K^r)$ as a function of $A$, for various values of $H_0^{}$. For $A = 0$ we recover the SdS situation and the value of $\theta\_E$ agrees with what one finds from Eq.\ (\ref{thetaE}). In particular, as we see from the plot and as can also be concluded more generally from Eq.\ (\ref{EuclideanAngle}), when we use our criterion for identifying light rays in different spacetimes, based on fixing the values of $r\_S$ and $r\_L$, the Euclidean angle in SdS spacetime does not depend on $H_0^{}$, contrary to what one would conclude by fixing $b$; when $A \ne 0$, however, $\theta\_E$ does depend on $H_0^{}$.

Figs.\ \ref{angleS} and \ref{angleC} show the more physically meaningful curved-geometry angle $\theta\_M$ measured by static and comoving observers, respectively, as functions of $A$ for various values of $H_0$. It can be checked that for $A = 0$, in SdS spacetime, the value of the static-observer angle $\theta\_M$ decreases with increasing $H_0$, in agreement with (\ref{thetastatic}). In order to study the slope of the measured bending angle versus acceleration ($A$) for a comoving observer, let's recalculate $\cos\theta_{\text{com}}$ from (\ref{thetao}), using the comoving observer velocity $U^\mu_{\rm com}$ in (\ref{4velo}) and the null radial vector $W^\mu$ in (\ref{radnull}),
\beq
\cos (\theta_{\text{com}})
= \frac{K^r-r\sqrt{f(r)}\, H(t) K^t}{f(r) K^t}
= \frac{1}{f(r)}\,\frac{K^r}{K^t} - \frac{r}{\sqrt{f(r)}}\,H(t)\;.
\eeq
If one further uses a linear polynomial as in (\ref{linear}) for $H(t)$, the variation of $\theta_{\text{com}}$ as a function of $A$, keeping $r = r\_L$ fixed, can be a calculated from
\beq
\frac{\partial}{\partial A}\cos(\theta_{\text{com}})
= \frac{1}{f(r)}\,\frac{\partial}{\partial A}\left(\frac{K^r}{K^t}\right)
- \frac{r}{\sqrt{f(r)}}\,t\;. \label{Acomslope}
\eeq
Since our observer is far from the central mass, $f(r) \approx 1$ and $K^r \approx K^t$, so the second term in (\ref{Acomslope}) dominates and the slope of $\theta_{\text{com}}$ vs. $A$ is positive, as seen in Fig. \ref{angleC}.


\section{Conclusions}
\label{Conclusions}
\noindent In this paper we examined the bending of null geodesics in spatially flat McVittie metrics, exact solutions to the Einstein equation which have been shown to represent nonrotating black holes embedded in FLRW background spacetimes. We used a slowly varying Hubble parameter $H(t) = H_0 + A\,(t-t_0)$, and the null geodesics were found numerically for situations in which the source and lens were aligned as seen by the observer. Simulations were run with fixed values for the mass of the central object and distances from the observer to the source and lens, in static coordinates, while we used three different values for $H_0^{}$ and in each case we looked at how the angle of arrival of null geodesics at the observer's location varied with $A$; in the $A = 0$ case, our results are in agreement with earlier perturbative calculations for light bending in Schwarzschild-de~Sitter spacetimes \cite{Rindler1,Rindler2,KLake}.

Each simulation gave us three different values for the angle of arrival. The Euclidean angle (calculated using an auxiliary, fictitious flat spatial metric and therefore not physically measurable, but nevertheless useful for comparison with previous work and as a check on the resuls) is independent of $H_0^{}$ if there is no acceleration, $A = 0$, but interestingly our simulations show a small dependence on $H_0^{}$ if one considers cases with $A \ne 0$. The two other types of angles are the ones that would be measured by static and comoving observers, respectively, and both show a dependence on $H_0^{}$ as well as on $A$. In this model, using the worldlines of comoving observers as the better approximations to those of physical observers, and the results from the simulations with $H_0 = 70$ km/s/Mpc as more representative of the actual expansion rate, we find that over the full range of values we used for $A$, and with our values for the distances and lens mass the bending angle varies by about 5\% around $20''$ or so, which agrees in order of magnitude with the results found by Kantowski et al.\ in the Swiss-cheese model \cite{Kantowski}. These values may seem encouraging, but unfortunately the effects we described would most likely be dwarfed by measurement uncertainties and departures of the metrics around actual galaxy clusters from the spherically symmetric ones used here to model them. We view this work as a first step in quantifying the effect of $\dot H$ on light bending and lensing in a useful way, and an improved lens modeling is one way in which the work will have to be extended before these bending effects can be meaningfully related to measurements.

Two other ways in which our approach can be improved are the fact that we fixed the values of the source and lens coordinates $r\_S$ and $r\_L$ as if they were directly measurable, and the fact that we considered only cases in which the source, lensing object and observer are aligned. Obtaining results beyond the latter limitation is essentially straightforward, although one will have to extend our criterion for identifying geodesics in different spacetimes to a more general setting. To replace the criterion based on values of $r$ by a more realistic one based on redshifts, one would have to use the relationship between redshifts and distances, which depends on the cosmological expansion history. Related to this is the fact that, although the simulations themselves and the calculations of the arrival angles are non-perturbative, fixing values of $r$ is equivalent to fixing those of actual spatial distances only to leading order in $m$; and using the same values of $H$ in different spacetimes at the time the geodesic leaves the source is equivalent to using the same values of $H$ when the geodesics arrive at the observer's location also only to leading order.

\end{document}